\documentclass[conference]{IEEEtran}
\IEEEoverridecommandlockouts
\usepackage{cite}
\usepackage{amsmath,amssymb,amsfonts}
\usepackage{algorithmic}
\usepackage{graphicx}
\usepackage{textcomp}
\usepackage{xcolor}
\def\BibTeX{{\rm B\kern-.05em{\sc i\kern-.025em b}\kern-.08em
    T\kern-.1667em\lower.7ex\hbox{E}\kern-.125emX}}

\begin{document}

\title{Adaptive Captioning with Emotional Cues: Supporting DHH and Neurodivergent Learners in STEM
}

\author{
\IEEEauthorblockN{Sunday David Ubur}
\IEEEauthorblockA{\textit{Department of Computer Science} \\
\textit{Virginia Tech} \\
Blacksburg, Virginia, USA \\
uburs@vt.edu}
\and
\IEEEauthorblockN{Eugenia Ha Rim Rho}
\IEEEauthorblockA{\textit{Department of Computer Science} \\
\textit{Virginia Tech} \\
Blacksburg, Virginia, USA \\
eugenia@vt.edu}
\and
\IEEEauthorblockN{Denis Gracanin}
\IEEEauthorblockA{\textit{Department of Computer Science} \\
\textit{Virginia Tech} \\
Blacksburg, Virginia, USA \\
gracanin@vt.edu}
}

\maketitle
\maketitle
\thispagestyle{plain}

\begin{center}
\footnotesize
This is the author's accepted manuscript of a paper presented at the
2025 13th International Conference on Affective Computing and Intelligent Interaction (ACII 2025).
The final version is to appear in IEEE Xplore. DOI will be added when available.
\end{center}

\begin{abstract}
Real-time captioning is vital for Deaf and Hard of Hearing (DHH) and neurodivergent learners (e.g., those with ADHD), yet it often omits emotional and non-verbal cues essential for comprehension.
This omission is particularly consequential in STEM education, where cognitively demanding material can exacerbate the challenges faced by caption users across diverse ability profiles.
In this paper, we present a design-oriented exploration of four captioning prototypes that embed emotional and multimodal cues, including facial expressions, body gestures, keyword highlighting, and emoji.
Across a pilot and a main study with 24 participants, we found that certain prototypes reduced self-reported cognitive load and improved comprehension scores compared to traditional captions.
Qualitative feedback reveals the importance of customizable caption features to accommodate neurodivergent users’ preferences (e.g., ADHD or different levels of comfort with emojis).

Our findings contribute to ongoing conversations in accessible technology research about how best to integrate emotional cues into captions in a way that is both usable and beneficial for a wide range of learners.
\end{abstract}


\section{Introduction}

Ensuring accessible learning in STEM (Science, Technology, Engineering, and Mathematics) remains a critical challenge for students who are Deaf or Hard of Hearing (DHH) as well as those who identify with neurodiversity, such as ADHD.
While automatic speech recognition (ASR) and Communication Access Realtime Translation (CART) services have improved, standard text-only captions can be cognitively taxing and still fail to convey vital non-verbal cues such as vocal tone, body gestures, and emotional emphasis~\cite{fink2020honoring,matsumoto2013facial,pang2024cross}.
This shortfall is especially problematic in STEM lectures, where emotional cues and prosodic emphasis signal difficulty or importance ~\cite{abu2019vocal,butler2018embodied, asee_peer_54067}.
Moreover, learners with conditions such as ADHD may particularly benefit from clearer, more dynamic forms of captioning that highlight emotional context and key phrases. ADHD students often struggle with sustained attention, benefiting from multimodal cues that emphasize key content \cite{10.1145/3579490, antonietti2021multimedia}.
By offering captions that provide not only textual content but also prosodic or facial indicators, we aim to support a broad spectrum of learning preferences and reduce extraneous cognitive load.

Over the past several years, there has been emerging work on ``emotive captions'' that embed or visualize paralinguistic speech cues, from pitch and volume to facial and gestural information~\cite{de2023visualization,kim2023visible,ohene2007emotional}.
Within the accessibility community, notable examples include research into ``Caption Royale,'' which explores a design space for affective captions~\cite{de2024caption}, ``Visible Nuance''~\cite{kim2023visible}, and others examining user-driven customization of emotional or prosodic cues in captioning systems~\cite{10.1145/3613904.3642177,10.1145/3613904.3642162}.
These studies consistently show that adding emotional or paralinguistic visualizations can improve user comprehension and engagement, but they also raise important questions about clutter, visual load, and user preferences.
Further, many of these projects focus on entertainment media, social platforms (e.g., TikTok, YouTube), or pre-recorded materials—leaving open questions about real-time usage, especially in formal educational contexts like STEM, where domain-specific complexity could introduce additional cognitive burden.

This work explores how to enrich real-time captioning with emotional and multimodal cues for STEM learners through a series of design probes. We present four prototypes that vary in the mode of emotional expression (e.g., on-screen emoji, text highlighting, embedded speaker video) and in how closely the captions are visually integrated with the instructor’s face or gestures. Specifically, we ask:

\begin{description}
\item[RQ1:]
How does the integration of facial expressions, text highlighting, emoji, or other emotional cues in real-time STEM captions affect comprehension and cognitive load for DHH and neurodivergent students?
\item[RQ2:]
In what ways do different approaches to visualizing emotional and prosodic information (e.g., on-screen emoji, overlay on the speaker video, textual emphasis) shape user preferences, perceived usability, and perceived connection with the instructor?
\end{description}

In a pilot study ($n=5$) and a subsequent main study ($n=24$), we evaluated user perceptions, cognitive load (via NASA-TLX~\cite{hart2006nasa}), and task comprehension across prototypes.
Although limited by a modest sample size and partial ordering effects, our initial findings suggest that certain ``emotion-enhanced'' designs can reduce participants' reported cognitive load compared to conventional captions and bolster comprehension on short quizzes.

\subsection{Contributions}

Our research makes the following contributions to the field of accessibility and education:

\begin{itemize}
\item
\textbf{Novel Emotive Captioning Approach:}
We introduce a real-time captioning framework that goes beyond standard text by integrating emotional and multimodal cues (e.g., facial expressions, prosodic hints).
This design targets STEM settings, where subtle emphasis and instructor affect are vital for comprehension.
\item
\textbf{Empirical Evidence in STEM Contexts:}
Through a pilot and main study, we demonstrate that augmenting captions with emotional and multimodal elements can reduce cognitive load and improve short-term understanding of complex material, revealing the potential for more sophisticated captioning solutions in demanding academic domains.
\item
\textbf{Emphasis on Customization:}
We highlight how individual user preferences—especially among DHH and neurodivergent learners—vary widely, making flexible customization essential. Our findings suggest that user-driven toggles for emotive features (e.g., emojis, highlighting) can strike a balance between clarity and richness of communication.
\end{itemize}

\section{Related Work}

We discuss emotional cues in education, emotional expressiveness and multi-modality for real-time captions.

\subsection{Emotional Cues in Education} 

Emotional expression in instructional settings involves nonverbal cues such as facial expressions, vocal tone, pitch, and body gestures.
These cues can convey an instructor's enthusiasm, emphasis, or urgency and are critical for effective communication—particularly in STEM classes where content is often dense and abstract~\cite{ekman1980relative,moeslund2011visual}.
While foundational research on universal emotions (e.g., anger, happiness) stresses the importance of nonverbal signals~\cite{ekman1992argument,mehrabian1981silent}, studies in educational psychology highlight that emotional cues can improve learner engagement and retention by signaling which topics are most important or conceptually difficult~\cite{sweller1994cognitive}.
In STEM lectures, instructors frequently rely on nonverbal behaviors to direct attention to key elements (e.g., pointing at equations or diagrams) and to communicate levels of confidence or urgency~\cite{walther2015nonverbal, 10.1007/978-3-031-60881-0_19}.
These cues are especially relevant for neurodivergent learners (e.g., those with ADHD or autism), who may process visual information more effectively than auditory cues~\cite{makhmudov2024enhancing}.
Research suggests that clarifying or translating these emotional and gestural elements into accessible formats---such as visually distinct text or icons in captions---can reduce cognitive load and help learners keep pace with complex material~\cite{bechtold2023cognitive,liu2022impacts, 10.1007/978-3-031-60881-0_24}.

\subsection{Emotional Expressiveness in Real-Time Captions}

Captions have traditionally focused on providing textual equivalents of spoken language for DHH communities.
More recently, research has expanded to include emotional and prosodic elements in captions to capture a fuller communication experience~\cite{ohene2007emotional,kim2023visible}.
For example, studies show that including cues for vocal intonation (e.g., indicating raised pitch or emphasis) can help readers follow important shifts in content or speaker intent~\cite{shigeno2016speaking, de2023visualization}. These approaches use font or animation to reflect prosody but must avoid overwhelming viewers~\cite{rashid2006expressing,malik2009communicating}.
In STEM education, emotional and prosodic cues can be especially valuable.
Lessons may involve complex terminology, rapid shifts in difficulty level, and specialized gestures (e.g., pointing at a chemical structure or highlighting code syntax).
Incorporating these cues into real-time captions can help DHH and neurodivergent learners identify moments of emphasis or conceptual change more quickly~\cite{brown2015dynamic,schlippe2020visualizing}.
However, existing solutions often target entertainment or general online content rather than highly technical, fast-paced STEM contexts~\cite{ohene2007emotional}.
Consequently, more research is required to understand how captioning systems can seamlessly integrate emotional cues, instructor gestures, and specialized vocabulary under stringent real-time constraints.

\subsection{Integrating Multiple Modalities in Real-Time Captions}

Recent work in multimodal captioning studies how to fuse textual information with visual data, such as facial expressions, voice pitch, and body gestures to enrich the accessibility experience~\cite{viegas-etal-2023-including, jia2014head}.
For instance, systems that analyze facial cues can display emotive icons or text effects indicating the speaker’s emotional state, while voice pitch analysis can inform dynamic text changes signaling urgency or enthusiasm~\cite{VILLEGASCH2025200479,tang2008eava}.
Although promising, many of these approaches focus on pre-recorded content and do not address the unique challenges of live, interactive lectures~\cite{li2021write, oshiba2023face}.
In the STEM classroom, capturing nonverbal cues in real time is further complicated by instructor movement around the room, the need to reference multiple visual aids (e.g., whiteboards, slides, lab equipment), and potential occlusions of facial expressions~\cite{grundmann2021face}. Some learners benefit from gesture icons, while others—especially with ADHD—may find them distracting~\cite{dollinger2021training}.
Thus, customizable captioning features—ranging from selectable emotional icons to on/off toggles for gesture annotations, could better accommodate neurodivergent preferences.
Overall, while multimodal and emotion-enhanced captioning has made strides in entertainment and general communication contexts, STEM education remains underexplored.
Prior solutions often do not address the specific demands of technical lectures that involve diagrams, formulas, and rapid topic shifts.
Nor have they sufficiently investigated how neurodivergent students perceive or utilize emotional cues in real-time captions.
This gap motivates our focus on a customizable, emotion-enhanced captioning system designed to maintain clarity while conveying crucial affective and instructional signals in STEM settings.

\section{Formative Pilot Study}

Before developing our main prototypes, we ran a brief pilot ($n=5$) to explore how users perceived different ways of representing emotional or multimodal information in captions.
While small in scale and without formal hypothesis testing, the pilot provided early signals that users want more than mere text but worry about clutter.

\subsection{Procedure and Materials}

\begin{figure}[htbp]
\centering
\includegraphics[width=\columnwidth]{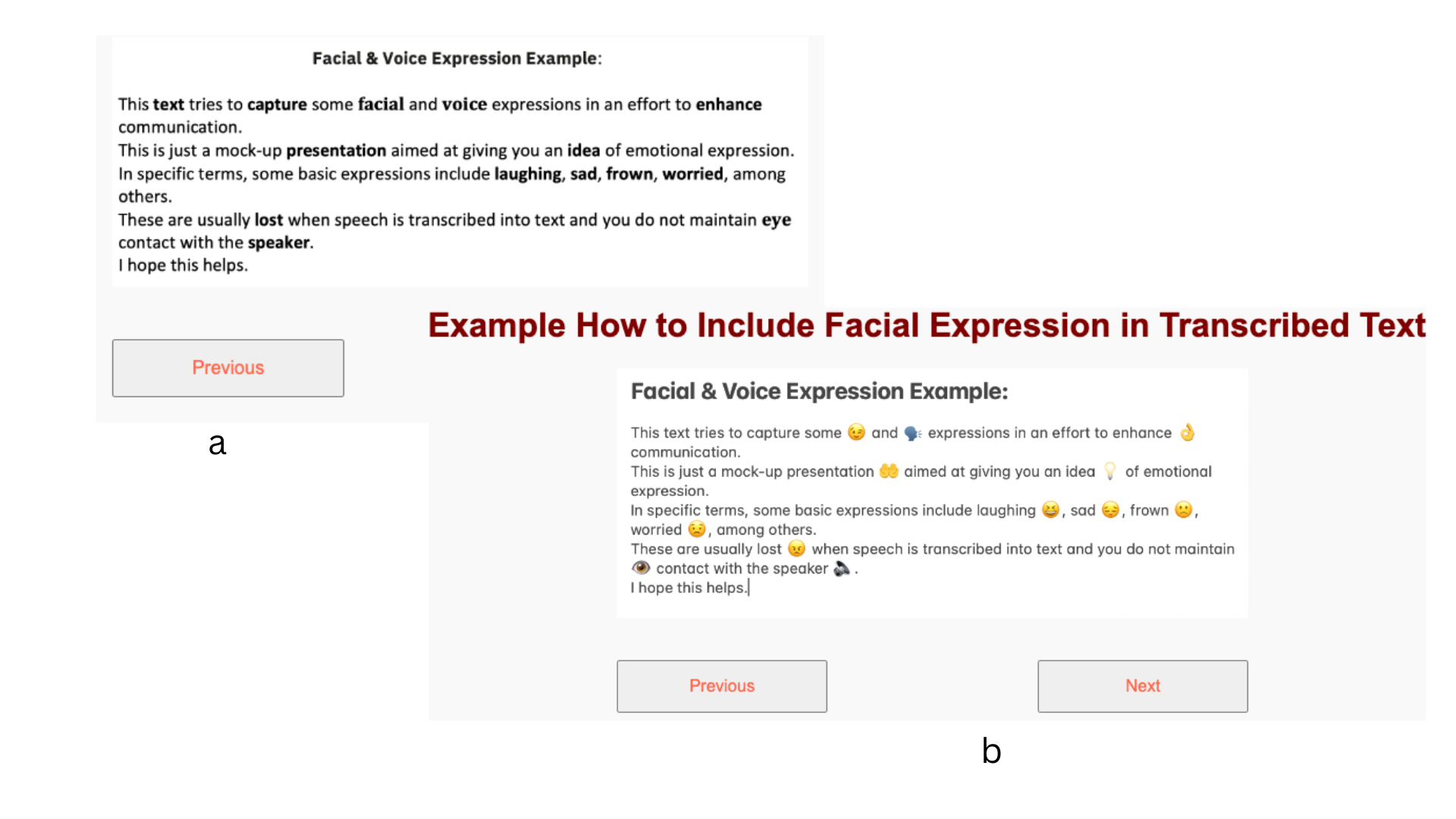}
\caption{Proposed caption design mock-up}
\label{fig:pilot_img1}
\end{figure}

Participants first watched a short recorded STEM lecture (Chemistry, 3~minutes) with only raw ASR transcripts.
They then reviewed design mockups illustrating possible enhancements:
(i) bold or colored highlighting of key/emotion-laden words, and
(ii) emojis inserted to convey speaker mood (Figure~\ref{fig:pilot_img1}.
We collected open-ended feedback via an online survey (QuestionPro).

\subsection{Pilot Feedback and Refinements}

Several found text-based highlighting helpful—one stated ``it could help me notice important points faster'' but were uncertain whether it conveyed genuine emotion.
Emojis received mixed reactions: some welcomed them as ``fun and quick to parse,'' while others felt they might be ``unprofessional or distracting for a formal STEM video.''
One Deaf participant said,
``I prefer reading standard English.
Emojis are okay in chats, but not sure in class.''
Despite these differences, participants unanimously agreed that a single ``one-size-fits-all'' approach was inadequate, suggesting that caption systems should allow customizations.
From these insights, we refined four prototypes, each representing a different blend of visual and emotional cues.
We then evaluated these prototypes in a main study with 24 participants.

\section{Main Study}

\subsection{Overview and Rationale}

Our main study aimed to compare a baseline traditional caption format with four enhanced prototypes, each embedding varying degrees of emotional expression or visual annotation.
We view this primarily as a design exploration rather than a confirmatory experiment, given the small sample and the multiple confounds (e.g., varying STEM topics).
Nonetheless, we sought indicative evidence on whether emotive captions might reduce cognitive load or improve short-term comprehension.

\subsection{Prototype Design}
\label{sec:prototype_design}

We developed a web-based captioning prototype that embeds emotional cues into STEM lectures, aiming to highlight nonverbal aspects (e.g., vocal emphasis, instructor gestures) for learners who are DHH or neurodivergent.
Implemented in HTML, CSS, and JavaScript, the simulated model processes pre-recorded ~4-minute lecture segments, segments the transcript into individual words, and annotates them with affective attributes such as color, animation, or icons.
This approach offers a more expressive form of real-time captioning, reducing cognitive load by visually conveying tone and urgency.

To explore diverse design choices, we created four discipline-specific variations, each featuring progressively richer capabilities:

\begin{itemize}
\item
\textbf{P1: Text + Speaker Video:}
Plain text captions appear to the side of the instructor’s video, enabling viewers to see facial expressions and gestures in real time.
Although unobtrusive, this layout contains no additional visual emphasis beyond sequential text.
\item
\textbf{P2: Caption Overlay on Video:}
Captions are overlaid on top of the speaker’s video, reducing the need to glance away.
However, some participants cautioned that occluding lip movements or facial expressions could hinder users who rely heavily on visual cues.
\item
\textbf{P3: Text Highlighting + Minimal Animation:}
Key terms (e.g., \emph{``electron,'' "important," "warning"}) are bolded or colored to indicate emphasis, with subtle text-rising animations simulating prosodic changes. This design avoids emojis, anticipating that some users might find them distracting.
\item
\textbf{P4: Text + Emojis/Icons + Discipline Visuals:}
Building on P3, this most visually "rich" prototype integrates small emojis or discipline-specific icons (e.g., lab flasks for chemistry) to reinforce emotional states (e.g., excitement) or specialized vocabulary. While potentially more engaging, it also introduces the highest risk of visual clutter.
\end{itemize}

Although each prototype relies on the same underlying textual transcript, we paired each with a short lecture from a different STEM domain (e.g., Engineering for P2, Biology for P3, Chemistry for P4) to reduce learning or memorization effects across conditions.
We acknowledge the potential for domain-based confounds (some subjects may inherently seem simpler or more engaging), but this variety allowed us to evaluate how emotional captioning can adapt to a range of technical complexities.

\subsection{Participants}

Through academic mailing lists and DHH advocacy groups, we recruited 33 participants, with 24 completing all study phases (72.7\% completion rate).
Gender distribution was balanced (54.2\% female, 33.3\% male, 12.5\% other/non-binary), with ages ranging from 19–40+ years.

\subsubsection{Accessibility Characteristics}

The participant cohort (Table \ref{tab:demographics}) included key populations of interest: 33.3\% (n=8) identified as Deaf or Hard of Hearing (5 Deaf, 3 Hard of Hearing), 29.2\% (n=7) reported ADHD, and 4.2\% (n=1) reported dyslexia. Caption usage patterns revealed that 92\% were regular users (split evenly between always and frequent users), and 58.3\% rated captions as "extremely useful" for their needs.

\begin{table}[htbp]
\caption{Participants Demographic Information, Accessibility Characteristics, and Caption Usage}
\centering
\scriptsize
\begin{tabular}{|c|c|c|c|c|c|}
\hline
\textbf{ID} & \textbf{Age} & \textbf{Gender} & \textbf{Profession} & \textbf{Accessibility} & \textbf{Caption Usage} \\
\hline
1 & 33 & Female & Industrial & ADHD & Always \\
2 & 33 & Female & Industrial & ADHD & Always \\
3 & 26 & Female & Industrial & Other & Frequently \\
4 & 40 & Female & STEM Student & Other & Frequently \\
5 & – & Female & STEM Student & DHH & Always \\
6 & 23 & Other & STEM Student & ADHD & Frequently \\
7 & 21 & Female & STEM Student & ADHD & Frequently \\
8 & – & Female & STEM Student & – & Always \\
9 & 23 & Other & STEM Student & ADHD + DHH & Frequently \\
10 & 38 & Other & Other & ADHD & Frequently \\
11 & 19 & Female & STEM Student & DHH & Always \\
12 & 32 & Female & STEM Student & ADHD + DHH & Frequently \\


13 & 21 & Female & Other & DHH & Frequently \\
14 & 28 & Male & Other & – & Always \\
15 & – & Male & Industrial & DHH & Occasionally \\
16 & 37 & Male & Other & DHH & Always \\
17 & 38 & Male & Other & – & Always \\
18 & 38 & Male & Industrial & – & Always \\
19 & – & Female & Other & – & Always \\
20 & 22 & Female & Industrial & Other & Frequently \\
21 & 22 & Female & Non-STEM & Other & Frequently \\
22 & 24 & Male & Non-STEM & Dyslexia & Always \\
23 & – & Male & Other & DHH & Frequently \\
24 & – & Male & Academic & – & Never \\
\hline
\end{tabular}
\label{tab:demographics}
\end{table}

\subsubsection{Rationale for Sample Composition}

This distribution enabled us to:
\begin{itemize}
\item
Compare DHH vs. hearing caption users' experiences.
\item
Examine ADHD-related preferences (given 29\% prevalence).
\item
Assess professional vs. student needs.
\end{itemize}

While the DHH subgroup (n=8) provided essential accessibility insights, we acknowledge this as a study limitation and recommend future work with larger DHH samples. The inclusion of frequent caption users (92\% of participants) ensured informed evaluations of prototype enhancements.

\subsection{Procedure}

\subsubsection{Consent and Pre-Questionnaire}

Gathered demographics, prior caption usage, and any accessibility needs.

\subsubsection{Part One (Baseline)}

Participants watched a short (~3 minute) lecture clip with traditional captions (raw text from ASR, no markup).
They answered two comprehension questions and completed the NASA-TLX~\cite{hart2006nasa} for cognitive load.

\subsubsection{Part Two (Enhanced Prototypes)}

Each participant then experienced the four prototypes in a randomized order (to mitigate learning or fatigue effects).
For each prototype, they: watched a short STEM clip (~4 minutes) with the prototype's enhancements, answered 2–3 comprehension questions, and completed NASA-TLX subscales.

\subsubsection{Post-Study Survey}

Gathered open-ended feedback on preferences and perceived benefits or drawbacks of each prototype.

\subsubsection{Note on Counterbalancing}

We acknowledge that using different video topics (Engineering, Biology, etc.) for each prototype can introduce content-based variability.
We also showed the baseline first, which does risk an order effect if participants ``learned how to answer'' by the time they reached the prototypes.
Because our focus was exploratory, we accepted these limitations but highlight them as a methodological caveat.

\subsection{Measures and Analysis}

\subsubsection{Comprehension Accuracy}

Binary scoring for each question (correct/incorrect).

\subsubsection{Cognitive Load}

We used NASA-TLX subscales (mental demand, frustration, etc.).

\subsubsection{Qualitative Feedback}

We performed a thematic analysis of open-ended survey responses, seeking insights on usability, emotional clarity, and perceived trade-offs.

We used repeated-measures ANOVA and $t$-tests to compare comprehension and NASA-TLX scores. Where relevant, we report effect sizes (partial $\eta^2$) to indicate practical significance. However, given the small sample and partial confounds, these results should be interpreted cautiously.

\section{Results}

\subsection{Comparing Baseline vs. Enhanced Prototypes}

We first compared how many participants answered comprehension questions correctly in the baseline vs. the four enhanced prototypes.
In the baseline condition, only 54.17\% of responses were correct overall.
In contrast, each enhanced prototype yielded markedly higher correctness rates—reaching nearly 100\% across participants.
An independent $t$-test showed that these differences were statistically significant ($p<.001$), albeit confounded by the fact that different lecture content was used.

\begin{table}[htbp]
\centering
\caption{Repeated-measures ANOVA results for NASA-TLX subscales. 
Values in parentheses represent partial $\eta^2$.}
\label{tab:anova1}
\setlength\tabcolsep{3pt}
\begin{tabular}{|l|c|c|c|}
\hline
\textbf{Subscale} & \textbf{F-value} & \textbf{p-value} & \textbf{partial $\eta^2$} \\ \hline
Mental Demand & 
  $F(3,10824.458)=4.6065$ & $p=0.0060$ & 0.0013 \\ \hline

Physical Demand &
  $F(3,9221.0348)=3.4898$ & $p=0.0216$ & 0.0011 \\ \hline

Effort &
  $F(3,8279.4334)=3.6820$ & $p=0.0173$ & 0.0013 \\ \hline

Frustration &
  $F(3,20878.496)=6.7620$ & $p=0.0006$ & 0.0010 \\ \hline
\end{tabular}
\end{table}

\subsection{NASA-TLX Results Across Prototypes}
\label{sec:nasa-tlx}

\begin{figure}[htbp]
\centering
\includegraphics[width=\columnwidth]{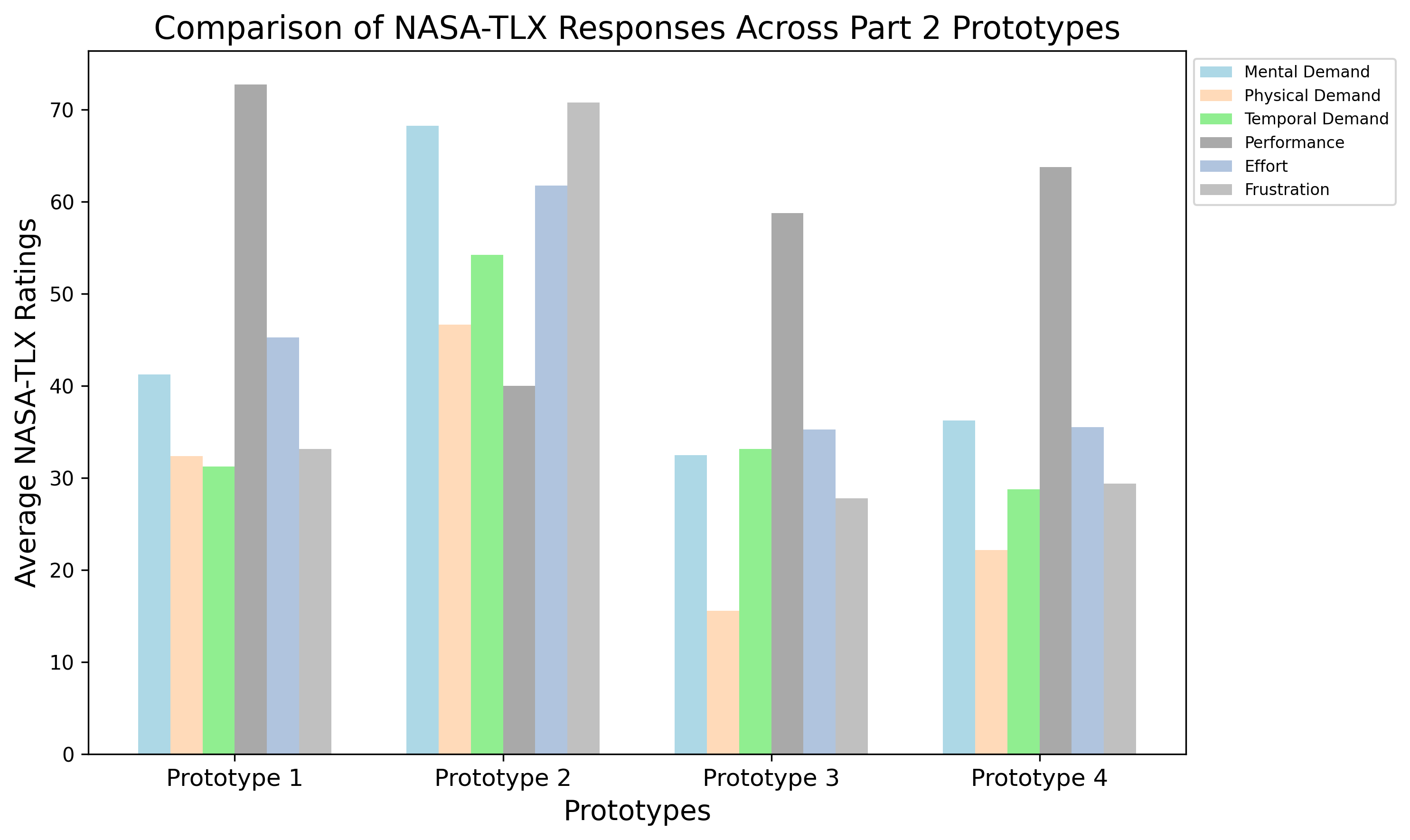} 
\caption{NASA-TLX scores for Prototypes 1--4.} 
\label{fig:nasa_tlx} 
\end{figure}

As shown in Figure~\ref{fig:nasa_tlx}, the baseline condition ($M=44.0$) and Prototype~2 ($M=68.3$) had relatively high Mental Demand ratings, whereas Prototype~3 was the lowest ($M=32.5$).
Participants noted that the overlaid captions in P2 frequently obscured the speaker’s mouth and gestures, posing additional challenges for those who rely on lip-reading or visual facial cues.
By contrast, \textit{Frustration} peaked under Prototype~4 (approximately $70$), while P3 elicited moderate frustration (approximately $50$).
\textit{Physical Demand} remained in the mid-range ($\approx 30$--$40$), and \textit{Temporal Demand} plus \textit{Performance} showed less pronounced variation.
\textit{Effort} roughly mirrored Mental Demand, with higher values for P2 ($\approx 60$) and lower for P3 ($\approx 40$).

A repeated-measures ANOVA (Table~\ref{tab:anova1}) revealed significant effects of prototype on \textit{Mental Demand}, \textit{Physical Demand}, \textit{Effort}, and \textit{Frustration} (all $p<0.05$).
Despite reaching statistical significance, effect sizes were small (partial $\eta^2 \approx 0.001$), likely reflecting the large error term in our model.
Specifically for Mental Demand, the effect was:
\[
F(3,10824.458)=4.6065\;\;\; p=0.0060\;\;\; \text{partial }\eta^2=0.0013
\]
Post-hoc comparisons confirmed that P2 was significantly more demanding than P3 ($p<0.05$).
Overall, these findings suggest that P2's overlaid design may impose additional visual load, while the minimal, keyword-highlight approach in P3 was comparatively less mentally taxing.

\subsection{Identification of Multimodal Cues and Effect}
\begin{figure}[htbp]
\centering
\includegraphics[width=\columnwidth]{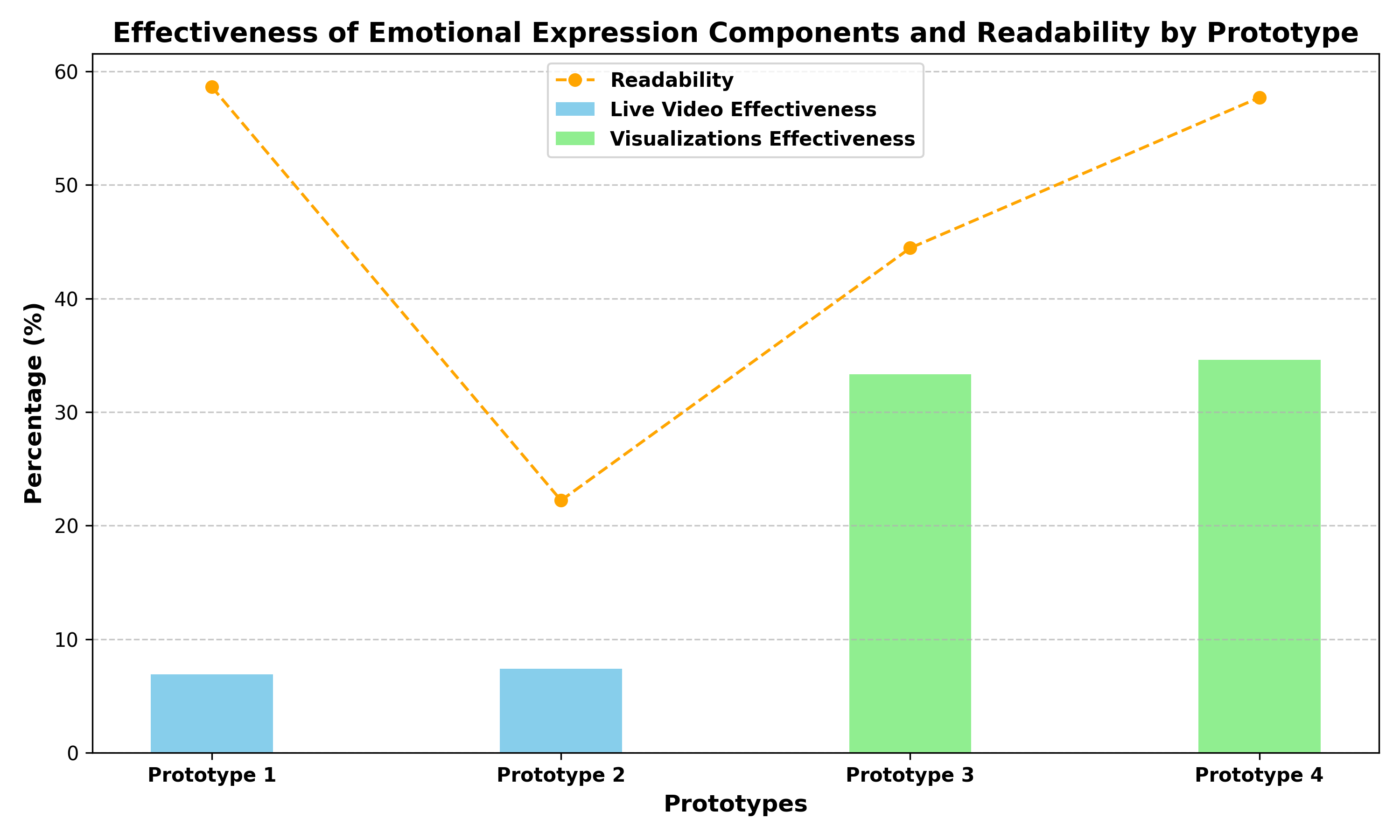}
\caption{Effectiveness of emotional expression components and readability across Prototypes 1--4.} 
\label{fig:rq1} 
\end{figure}

Figure \ref{fig:rq1} describes the effectiveness of emotional expression components and readability across the four prototypes.
Live video (Prototypes 1 and 2) showed limited effectiveness in conveying emotions, with only 6.90\% and 7.41\% of users finding it effective, respectively.
In contrast, visualizations and text effects (Prototypes 3 and 4) were more effective, with 33.33\% and 34.62\% of users identifying emotions successfully.
Readability was highest for Prototypes 1 and 4, scoring 58.62\% and 57.69\%, while Prototype 2 had the lowest readability at 22.22\%.

\subsection{Captioning Preferences and Usage Trends Across Respondent Groups}

\begin{figure}[htbp] 
\includegraphics[width=\columnwidth]{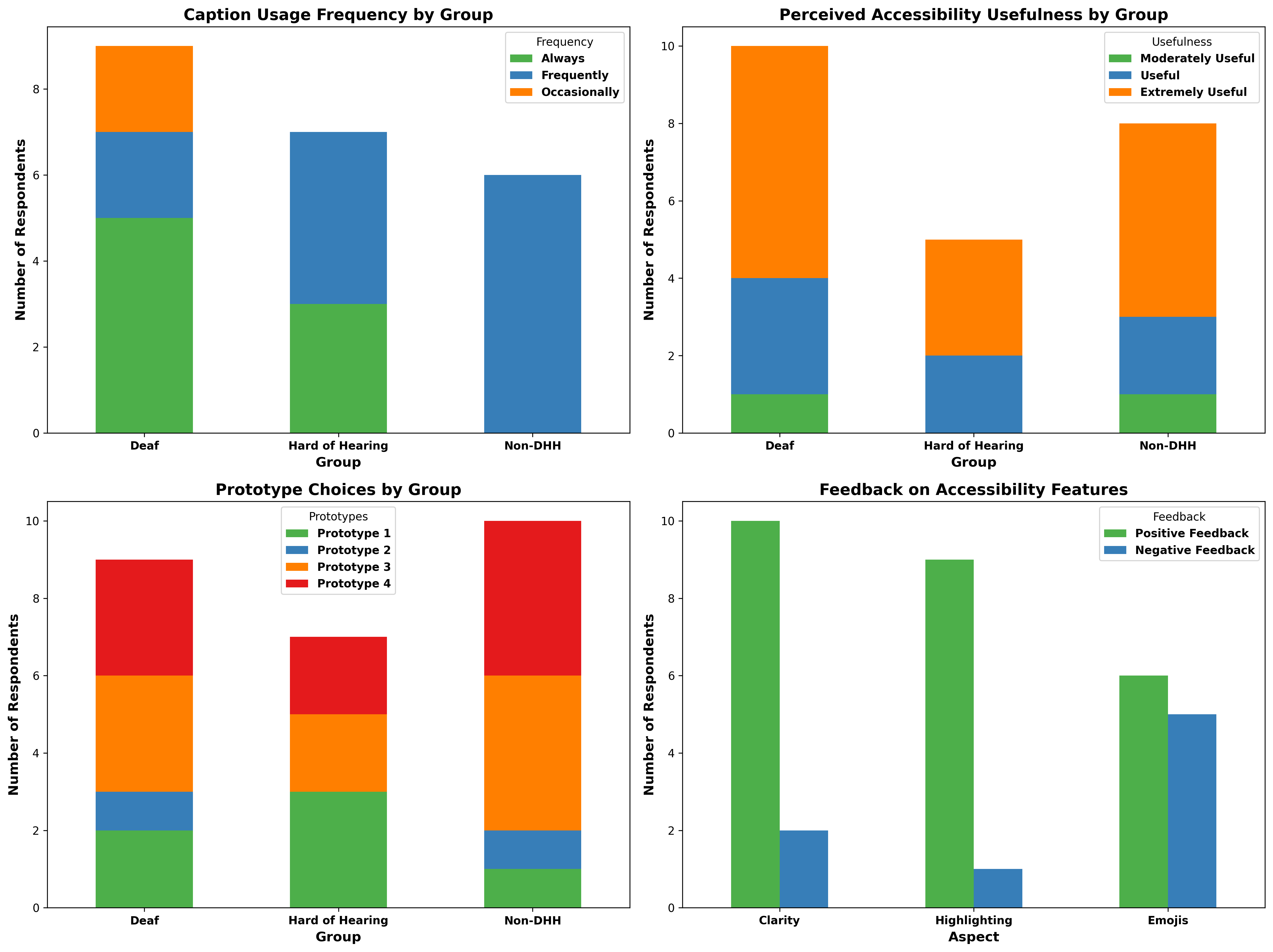}
\caption{
Comparison of caption usage frequency, perceived usefulness, prototype preferences, and feedback on accessibility features across Deaf, Hard of Hearing, and neurodivergent respondents.
} 
\label{fig:preferences} 
\end{figure}

Figure~\ref{fig:preferences} show clear differences in how captions are used and which prototypes are preferred by Deaf, Hard of Hearing (HoH), and neurodivergent respondents.
Deaf participants mostly selected captions as ``Always'' and leaned towards Prototypes 3 and 4 because they effectively conveyed emotional tone and context, although they pointed out that excessive distractions, like emojis, were a drawback.
HoH respondents valued simplicity, with Prototype 1 receiving significant praise for its clarity and the way it separates text from video.
neurodivergent participants liked features such as keyword highlighting in Prototype 3 and found Prototype 4 easy to read, often mentioning accessibility and cognitive clarity as important factors.
Overall, all groups emphasized the importance of simplicity, clarity, and minimizing distractions for perceived usefulness and accessibility.

\subsection{Qualitative Perceptions}

In open-ended feedback, participants varied considerably on the value of emoji or domain-specific icons (P4).
Some wrote, ``Emojis helped me see if the instructor was excited or joking,'' while others described them as ``unnecessary fluff.''
Two Deaf participants specifically mentioned that emojis felt distracting and that ``clean text with color or bold is better than cartoony icons.''
Meanwhile, P3—highlighting key words without emoji—was praised for balancing expressiveness and legibility.
One ADHD participant noted, ``Highlights drew my attention to important points without feeling overwhelmed.''

Interestingly, some participants still favored P1 (separate caption box plus full speaker video), citing that ``it’s easiest to read in one place while still seeing the face.''
Another participant, though, disliked having to shift gaze from text to video.
These tensions suggest personalization is crucial: for instance, letting each user choose overlay vs. side-by-side text, or toggling off emoji.

\subsection{Summary of Findings}

\subsubsection{Prototype Variation}

Not all enhancements uniformly helped.
Certain design choices (like text over the speaker) was frustrating for some, and emoji were polarizing.

\subsubsection{Reduced Cognitive Load}

Relative to the baseline, participants generally reported reduced cognitive effort with prototypes that integrated emotional cues in a subtle, text-centric way (P3).

\subsubsection{Strong Demand for Customization}

Over half of participants explicitly requested ways to toggle or personalize features.

\subsubsection{Preliminary Evidence of Improved Comprehension}

Although partly confounded by different content, participants did better on short quizzes with emotive or visually emphasized captions.

\section{Discussion}

Our study set out to examine how embedding emotional and multimodal cues into real-time captions might improve comprehension and reduce cognitive load for STEM learners. Specifically, we asked:

\begin{description}
\item[RQ1:]
How does the integration of facial expressions, text highlighting, emoji, or other emotional cues in real-time STEM captions affect comprehension and cognitive load for DHH and neurodivergent students?
\item[RQ2:]
In what ways do different approaches to visualizing emotional and prosodic information shape user preferences, perceived usability, and perceived connection with the instructor?
\end{description}

\subsubsection{Addressing RQ1}

Our findings suggest that including emotional indicators—such as highlighting of keywords, emojis, and minimal text animation—can reduce perceived mental effort and increase short-term comprehension.
The participant group that experienced these enhancements showed a jump from 54.17\% correctness with traditional captions to nearly 100\% correctness across the four prototypes.
While some of this improvement is likely attributable to practice effects or differing lecture content, the self-reported NASA-TLX measures also revealed that prototypes incorporating subtle emotional cues (e.g., color highlighting) were perceived to be less mentally demanding than purely text-based captions.
These results align with theories in cognitive load, which posit that well-designed multimodal cues can lower extraneous processing and support deeper engagement with STEM content~\cite{bechtold2023cognitive,liu2022impacts}.

\subsubsection{Addressing RQ2}

Though each prototype included a distinct configuration of emotional cues, user comments underscored how varied preferences can be.
Prototypes that inserted emojis received polarized feedback, praised by some for clarity and immediacy (``I could see how the speaker was feeling'') but criticized by others as ``cluttered'' or ``distracting.''
Overlaid captions (Prototype 2) simultaneously allowed for eye-contact alignment but also frustrated participants who rely on lipreading.
By contrast, many users lauded the simpler text-highlighting approach (Prototype 3), which struck a balance between expressiveness and readability.
These findings highlight the challenge of conveying emotional cues without visual overload.~\cite{kim2023visible,ohene2007emotional}. A single method for emotional expression does not appear optimal across diverse needs; customization and adaptability emerged as recurring themes in our data.

\subsection{Methodological Limitations}

Despite these promising indicators, the exploratory nature of our study introduces several limitations that constrain the interpretability of our results.

\subsubsection{Study Population and Sample Size}

We recruited 24 participants, among whom only 8 identified as DHH.
While this subset allowed for some initial insights into the unique needs of DHH learners, it is insufficient for conclusive, generalizable claims about the entire DHH population.
Future studies should broaden recruitment efforts to capture a wider range of DHH users, including those with varying levels of fluency in sign language, different familiarity with captioning, or additional needs such as cochlear implants.

\subsubsection{Potential Order and Learning Effects}

We chose to show the baseline (traditional captioning) condition first, then randomized only the order of the four enhanced prototypes.
Although this approach partially mitigates carry-over effects across the prototypes themselves, participants likely learned the basic content presentation format from the baseline condition before seeing the enhanced versions.
Moreover, each prototype was demonstrated with a different STEM topic, which introduces further content variability: some topics might have been inherently simpler or more engaging.
These design decisions were pragmatic for an exploratory pilot but limit our ability to make definitive causal inferences.

\subsubsection{Short Videos and Immediate Comprehension}

Our study used short STEM lectures (3–4 minutes) followed by a small number of factual or conceptual questions.
Participants' improved performance in prototypes may not necessarily extrapolate to longer lectures or more complex tasks involving multi-step reasoning or problem-solving.
Additionally, we relied on immediate comprehension checks, not longer-term retention or conceptual depth.
Future research should incorporate extended tasks (e.g., multi-lecture modules, multi-day classes) and delayed post-tests to better assess whether emotional cues in captions yield sustained learning benefits.

\subsection{Design Implications}

Our results suggest both opportunities and constraints when designing emotion-enhanced captioning systems.
Below, we synthesize key takeaways for practitioners and researchers.

\begin{itemize}
\item
\textbf{Subtle Emotive Cues May Reduce Cognitive Load:}
Previous research has emphasized that designers should balance expressiveness with clarity, ensuring that emotional cues supplement rather than dominate the visual field~\cite{rashid2006expressing,malik2009communicating}. Our exploratory study reinforces this principle in the underexplored context of real-time STEM captioning. Participants favored prototypes that used subtle visual enhancements—such as keyword highlighting without emojis—while describing overloaded designs with icons or animation as distracting. This suggests that even lightweight cues can enhance comprehension without adding unnecessary visual strain.

\item
\textbf{Avoid Overly Obstructive Overlays:}
Although some users appreciated caption-video overlays for visual alignment, others—especially those who rely on lipreading—found them intrusive. Prior work has raised similar concerns about overlays occluding facial features~\cite{brown2015dynamic,schlippe2020visualizing}. Our results underscore the importance of flexible layouts that preserve visibility of speaker expressions.

\item
\textbf{Customization is Critical:}
Prior work has noted that some Deaf users perceive emojis as ``childish'' or visually distracting~\cite{10.1145/3613904.3642177, 10.1145/3613904.3642162}. Our findings confirm this sentiment among several DHH participants, but also highlight a contrast: many participants with ADHD reported that emojis helped them quickly track emotional tone and stay engaged. This divergence underscores the need for user-driven customization—allowing individuals to enable, disable, or tailor emotive features like icons, highlights, or text animation based on personal preference.

\item
\textbf{Extended Accessibility Beyond DHH:}
While designed with DHH learners in mind, our findings reveal broader benefits. Neurodivergent users, including those with ADHD, and non-native English speakers reported improved focus and emotional clarity from enhanced captions. These cross-cutting benefits point toward inclusive design opportunities.
\end{itemize}

\subsubsection{Linking Back to STEM Pedagogy}

Emotional cues may be particularly beneficial in STEM contexts, where the instructor’s enthusiasm or caution can signal the difficulty or importance of a given concept~\cite{abu2019vocal,butler2018embodied}.
For instance, the intensity of a professor’s voice or a physical gesture might highlight a critical lab safety warning or a key theorem.
By translating these nonverbal expressions into visual cues—such as color shifts, bold formatting, or timely icons—learners can quickly identify high-stakes material or areas needing extra attention.
This alignment with STEM pedagogy underscores the unique value of emotive captioning in domains where missing subtle cues could have especially significant consequences, whether it involves complex calculus proofs or safety-critical lab procedures~\cite{fink2020honoring}.

\subsubsection{Connection to Accessibility Standards}

Given the broader push for inclusive design, emotive captioning aligns with existing accessibility guidelines such as WCAG and Section~508~\cite{w3cWCAG21}.
While these guidelines predominantly focus on text accuracy, proper synchronization, and navigability, our findings suggest that emotional expressiveness constitutes another layer of ``equivalent context,'' particularly in educational settings where tone and emphasis can shape understanding~\cite{pang2024cross}.
Researchers and practitioners could frame emotive captions as a complementary enhancement to meet or exceed the standards outlined in Deaf education best practices, thereby contributing to more holistic representations of speech.

\subsection{Toward Deeper Evaluation}
Given the exploratory nature of this study, future work should involve more rigorous, longitudinal investigations in real classroom environments.

\subsubsection{Long-Term Classroom Deployments}
Rather than short, self-contained lecture clips, future studies could instrument full course sessions over several weeks. This would enable analysis of how emotive captions affect not only immediate comprehension but also academic outcomes like homework, exams, and projects. Longitudinal use may reveal factors such as fatigue, novelty effects, or shifting preferences.

\subsubsection{Adaptive and Intelligent Systems}
Building on the demand for personalization, AI-driven captions could adapt emotive features based on user feedback. For instance, the system might reduce distracting emojis or amplify features like highlighting for certain learners. Real-time detection of speaker mood could dynamically adjust visuals to match vocal prosody.

\section{Conclusion}
We presented a design exploration of four prototypes integrating emotional expression and multimodal cues into real-time captions for STEM lectures.
Our pilot and main study indicate that certain enhancements can reduce self-reported cognitive load and boost short-term comprehension compared to conventional captions.
However, these findings also highlight notable disagreements among users regarding emoji and overlay text, reflecting diverse sensory and cognitive needs.

Despite design and sample limitations, this work contributes to ongoing discussions about enhancing captioning systems to be more expressive, accessible, and ultimately more reflective of the nuances in natural speech. We call for further research to rigorously test and refine emotive captioning approaches in real classroom or workplace environments, extending customizability and examining long-term educational outcomes. By embracing emotional cues in captions—and by giving users the freedom to adapt these cues—future systems can better serve the diversity of learners, particularly in the demanding context of STEM education.

\clearpage
\section*{Ethical Impact Statement}

\noindent
\textbf{Human Subjects and IRB Approval:} 
This research protocol was reviewed and approved by an Institutional Review Board (IRB \#24-612), meeting the criteria for exemption under 45 CFR 46.104(d).
All participants provided informed consent before taking part in the study.
Prior to data collection, each participant was informed of their right to withdraw at any time without penalty.
Personal identifiers were removed from data, and all results are reported in aggregated form to protect participant anonymity.

\vspace{2mm}
\noindent
\textbf{Recruitment, Compensation, and Demographics:}
Participants were recruited via a public email flyer and mailing lists catering to Deaf/Hard of Hearing (DHH) communities and individuals with neurodivergent conditions such as ADHD or autism.
The study took approximately 10 minutes.
In lieu of monetary payment for every participant, they were offered the option to enter a random drawing for one of five \$10 (USD) gift cards, drawn after the study concluded.
This approach was standardized for all participants, independent of performance or personal characteristics.
A total of 24 participants ultimately completed the study.

\vspace{2mm}
\noindent
\textbf{Privacy and Data Management:} 
All data—including survey responses and transcribed text logs—were stored on secure, encrypted servers accessible only to the research team.
Participant IDs were anonymized, and no personal identifiers (e.g., names, emails) were retained in the final dataset.
Data will be retained for up to three years for possible replication or follow-up analysis, after which it will be deleted or irreversibly de-identified.

\vspace{2mm}
\noindent
\textbf{Risk Assessment and Mitigation:} 
Participants were required only to watch short captioned videos and fill out a brief questionnaire.
There was no sensitive or invasive data collection.
The potential risk was minimal, primarily relating to privacy.
No covert or non-consensual sensing was used; participants could discontinue at any point without penalty.
All participants confirmed consent by reviewing a consent form detailing study goals, procedures, and voluntary participation.

\vspace{2mm}
\noindent
\textbf{Potential Biases and Fairness:}
We recognize that emotion-oriented captions or detection models could inadvertently reflect biases if trained on datasets not representative of diverse user groups, including signers or people with atypical facial movements.
Although this study did not deploy a large-scale AI model, we emphasize the importance of equitable data collection and transparent algorithms in future implementations.
Incorporating user-driven customization can help mitigate misclassification or marginalization.

\vspace{2mm}
\noindent
\textbf{Long-term Societal Impact:} 
Captioning systems integrating emotional and multimodal cues could substantially assist DHH and neurodivergent individuals, particularly in STEM contexts requiring high cognitive load.
However, misuse—such as unauthorized emotional tracking—remains a concern.
We advocate for strong consent frameworks, privacy protections, and responsible use guidelines to ensure such systems promote accessibility without facilitating involuntary surveillance.
If deployed responsibly, we believe these technologies can enhance communication equity across a broad range of learners.


\bibliographystyle{IEEEtran}
\bibliography{references}

\end{document}